\documentclass[preprint,12pt]{elsarticle}
\usepackage{amsmath}
\usepackage{graphicx}% Include figure files
\usepackage{dcolumn}% Align table columns on decimal point
\usepackage{bm}% bold math
\usepackage{caption}
\usepackage{xcolor}
\usepackage{float}
 \RequirePackage{mathptmx} 
\journal{Chinese Journal of Physics}

\begin{document}

\begin{frontmatter}

\title{Re-analysis of the Gamow-Teller distributions for  $N$=$Z$ nuclei, $^{24}$Mg, $^{28}$Si, and $^{32}$S}
\author{{Jameel-Un Nabi$^{a}$, Abdul Kabir$^{b}$ and Tuncay Bayram$^{c}$.}}
\address{$^{a}${University of Wah, Quaid Avenue, Wah Cantt 47040, Punjab, Pakistan.}}
\address{$^{b}${Space and Astrophysics Research Lab, National Centre of GIS and Space Applications, Department of Space Science, Institute of Space Technology, Islamabad and 44000, Pakistan.}}
\address{$^{c}${Department of Physics, Karadeniz Technical University, Trabzon, {61080}, Türkiye.}}

                    %% ABSTRACTT %%   
\begin{abstract}
The Gamow-Teller (GT) strength distributions of $sd$-shell $N$=$Z$ nuclei ($^{24}$Mg, $^{28}$Si, and $^{32}$S) are investigated within the framework of proton-neutron quasi-particle random phase approximation (pn-QRPA)  {using} a deformed basis. The nuclear properties of these special nuclei were investigated by the Relativistic Mean Field (RMF) model. The RMF framework with density-dependent interactions (DD-PC1 and DD-ME2) was employed to compute ground-state deformation parameters ($\beta_2$) using potential energy curves. The $\beta_2$ values computed from the RMF and the finite range droplet model (FRDM) were employed as a free parameter in the pn-QRPA calculation to investigate the resulting GT strength distributions. The present model-based analyses compare well with the observed data.

\end{abstract}
\end{frontmatter}
\section{Introduction}
\label{intro}
The analysis of Gamow-Teller (GT) strength of nuclei is an area of intense research and finds its application both in nuclear and astrophysics~\cite{Osterfeld}. 
The GT response is astrophysically relevant for a variety of nuclides, particularly those near iron (Fe). It contributes significantly to the weak-interaction rates~\cite{Bethe}.

{GT distributions were derived experimentally using multiple methods \cite{Anderson,Kateb,Fujita,Baumer,Cole}. Theoretical calculations of GT transitions are broadly classified into three types: simple independent-particle models; full-scale interacting shell-model calculations; and, in between, the random-phase approximation (RPA) and quasi-particle random-phase approximation (QRPA). The QRPA approach can be formulated based on the mean-field basis employing varying forms of potential. Noticeable mentions would include deformed Nilsson model \cite{Krumlinde,Staudt,Hirsch}, the finite-range droplet
model with a folded Yukawa single-particle potential \cite{Moller90,Moller97},
and the Woods–Saxon potential \cite{Hektor,Ni}. The QRPA technique has also been developed within the framework of finite Fermi system theory \cite{Krmpotic}. The residual interactions used in these QRPA techniques were not derived directly from the interactions essential to obtain the mean-field basis. It has been found that the
self-consistency of the QRPA approach is crucial in describing the GT strength function \cite{Engel3,Paar,Borzov,Borzov1}.}

{Reliable theoretical calculations of $\beta$-decay rates are difficult to obtain. Majority of the strength linked with the GT operator ($\vec{\sigma}$$\tau_{\mp}$) lies in GT resonance, well above the decay threshold.  The small low-energy tail of the GT distribution is the strength that genuinely plays a role in $\beta$-decay. This means that computed $\beta$-decay rates may be varied across a large range without breaking sum rules, which helps lower theoretical uncertainty for other processes. Preliminary attempts for a completely self-consistent computation were made in Ref.~\cite{Engel3}. They calculated $\beta$-decay rates in  even-even semi-magic nuclei lying on the $r$-process path.} 
{Martini \textit{et al.} \cite{Martini1} presented the extension of fully consistent QRPA  approach from electromagnetic to the charge-exchange nuclear excitations. They focused on the  role of nuclear deformation on the GT excitations and $\beta$-decay halflives. Zhiheng \textit{et al.} \cite{Zhihengw} employed the framework of the random-phase approximation (RPA) based on the relativistic Hartree-Fock (RHF) theory to achieve a self-consistent calculation with the $\rho$-meson tensor coupling. They investigated properties of the GT resonances.} 

The importance of the GT strength distribution in $sd$-shell nuclei has attracted the attention of  researchers due to recent developments in nuclear structure~\cite{Anderson,Grewe,Zegers}. GT transitions involve spin and isospin exchange, and mapping such strength, within a nucleus, serves as a stringent test of structure calculations~\cite{Brown}. The nuclei in the $sd$-shell are excellent {tools} for studying GT strength distributions. Anderson and collaborators \cite{Anderson} studied the charge-changing transitions in the $(p,n)$ reaction at 136~MeV on the self-conjugate $sd$-shell nuclei.  For each reaction, they analyzed the GT strength within $E_\mathrm{x}$=12 MeV in daughter nucleus. Grewe \textit{et al.,}~\cite{Grewe} studied the  $^{32}$S($d$, $^{2}$He)$^{32}$P reaction at  $E_d$=170 MeV. They extracted the B(GT$_\pm$) strength functions from the ($d$, $^{2}$He) reaction within $E_\mathrm{x}$= 7.01 MeV. Zeger \textit{et al.,}~\cite{Zegers} studied the $^{24}$Mg($^3$He, $t$)$^{24}$Al charge-exchange reaction  at {projectile incident energy of 420 MeV}. They extracted the GT strengths to discrete levels in $^{24}$Al up to 6.88 MeV. At zero momentum transfer, they used the empirical formulation for the proportionality of the GT strengths  and differential cross-sections. The more recently, Eunja \textit{et al.}  \cite{Eunja} have investigated GT strength distributions of $^{24}$Mg, $^{28}$Si, and $^{32}$S, employing a {deformed quasi-particle random phase approximation (DQRPA) model}. The DQRPA model used a realistic two-body interaction defined by the Brueckner
G-matrix based on the CD Bonn potential. For single-particle energies, they used the deformed Woods-Saxon potential and included the particle model space up to $fp$-shell. Additionally, they took into account deformation as well as connections between like- and unlike-pairing correlations.  They reported their GT strength estimates in relation to the parent nucleus excitation energy.

{In this project, we examine how nuclear deformation affects the predicted GT strength distributions for $N$=$Z$ nuclei.  We used the RMF model, involving two types of interactions, to compute two sets of nuclear deformation parameters. The RMF framework~\cite{Walecka} in the axially deformed shape with  density-dependent interactions was employed for the investigation of GT strength of nuclei: $^{24}$Mg, $^{28}$Si and $^{32}$S that have $N$=$Z$.} The RMF framework is a phenomenological method for the description of different ground-state nuclear characteristics of nuclei. {Further details of the model may be found in Refs.~\cite{lalazissis99,bayram13,Nabi}.} Additionally, {the RMF model provides information} regarding nuclei with hadronic degrees of freedom~\cite{Walecka}.
{In addition, we used the computed values of $\beta_2$ by FRDM framework~ \cite{Moller}.}
{The quadrupole deformation parameter $\beta_2$, using the FRDM and RMF models, were employed as a free parameter in the pn-QRPA model~\cite{Klapdor} to calculate GT strength for $sd$-shell nuclei with $N$=$Z$ ($^{24}$Mg, $^{28}$Si, and $^{32}$S).}

We have structured the paper so that section 2 briefly covers the requisite formalism. Section 3 discusses the findings of our investigation. Section 4 brings our results to few conclusions.
\section{Formalism}
\subsection{The pn-QRPA Model}
{The $\beta$-decay properties of the selected $sd$-shell nuclei were studied within the pn-QRPA approach using quasi-particle states obtained within a BCS + Nilsson model, with a separable (single plus quadrupole) multi-shell schematic interaction.}
The Hamiltonian of the pn-QRPA framework was chosen as
\begin{equation} \label{H}
	\centering
	H = H^{sp} + V^{pair} + V^{pp}_{GT} + V^{ph}_{GT},
\end{equation}
where $H^{sp}$, $V^{pair}$, $V_{GT}^{pp}$ and $V_{GT}^{ph}$ represent, the single-particle Hamiltonian, pairing potential, the particle-particle ($p$-$p$) and the particle-hole ($p$-$h$) interactions, respectively. The pairing force $V^{pair}$ was estimated using the BCS approximation.  {For the $Q$-value calculation, the recent mass compilation \cite{Wan21} was employed. The single-particle wavefunctions and energies were determined employing the deformed Nilsson framework \cite{Nilsson}. The oscillator constant for nucleons  was chosen as $\hbar\omega= 45 A^{-1/3}-25 A^{-2/3}$ MeV. The Nilsson potential parameters were taken from Ref.~\cite{Ragnarsson}.  Two different values of $\beta_2$ were used in the Nilsson framework. One was calculated from  the RMF framework (explained in the succeeding subsection) and the second was adopted from the FRDM calculation~\cite{Moller}.}

The transformation from the spherical basis of nucleon ($c_{jm}$, $c^{\dagger}_{jm}$) to the deformed nucleon basis ($d_{m\alpha}$, $d^{\dagger}_{m\alpha}$) was achieved by means of the transformation equation
\begin{equation}\label{df}
	\centering
	d^{\dagger}_{m\alpha}=\Sigma_{j}D^{m\alpha}_{j}c^{\dagger}_{jm},
\end{equation}
where $c^{\dagger}$ is the particle-creation operator in the spherical basis while $d^{\dagger}$ is in the deformed basis. {The matrices $D^{m\alpha}_{j}$ were determined from the diagonalization of the Nilsson Hamiltonian.  We assumed a constant pairing force with the interaction strength $G$ ($G_p$ and $G_n$ for protons and neutrons, respectively)}
\begin{equation}\label{pr}
	V_{pair}=-G\sum_{jmj^{'}m^{'}}(-1)^{l-m+j}c^{\dagger}_{j-m}c^{\dagger}_{jm}
		\\
		\times(-1)^{j^{'}-m^{'}+l^{'}} c_{j^{'}-m^{'}}c_{j^{'}m^{'}}.
\end{equation}
{The interaction strengths were fixed to the pairing energy gaps $\bigtriangleup_{pp}$ and $\bigtriangleup_{nn}$ for the proton and neutron systems.  These two latter quantities were computed using}
\begin{eqnarray}
	\begin{aligned}
		\bigtriangleup_{nn} =\frac{1}{8}(-1)^{A-Z+1}[2S_n(A+1, Z)- 4S_n(A, Z) \\
		+ 2S_n(A-1, Z)]
	\end{aligned}
\end{eqnarray}
\begin{eqnarray}
	\begin{aligned}
		\bigtriangleup_{pp} =\frac{1}{8}(-1)^{1+Z}[2S_p(A+1, Z+1)-4S_p(A, Z)\\
		+2S_p(A-1, Z-1)].
	\end{aligned}
\end{eqnarray}
{The BCS equation was then solved using these constant pairing forces. The correlations for neutron-neutron pairing and proton-proton pairing, which have solely isovector contributions, were included in the present analysis. For the isoscalar part, one has to include the neutron-proton pairing correlation, which is not considered in the present manuscript. The existing pn-QRPA model has the limitation of disregarding the neutron-proton pairing effect. The isoscalar  interaction behaves in a fashion similar to the tensor force interaction \cite{Ha22}. The tensor force shifts the GT peaks to low excitation energies. Incorporation of tensor force may result in lower centroid values of computed GT strength and could lead to higher values of calculated $\beta$-decay rates. The same effect of shifting calculated  GT strength to lower excitation energies in the current pn-QRPA model was achieved by incorporation of $p$-$p$ forces~\cite{Hirsch}.}

Later, a quasiparticle basis $(a^{\dagger}_{m\alpha}, a_{m\alpha})$ was presented using the Bogoliubov transformation
\begin{equation}\label{qbas}
	\centering
	a^{\dagger}_{m\alpha}=u_{m\alpha}d^{\dagger}_{m\alpha}-v_{m\alpha}d_{\bar{m}\alpha}
\end{equation}
\begin{equation}
	\centering
	a^{\dagger}_{\bar{m}\alpha}=u_{m\alpha}d^{\dagger}_{\bar{m}\alpha}+v_{m\alpha}d_{m\alpha},
\end{equation}
{where $\bar{m}$ is the time-reversed state of $m$ (third component of the angular momentum), $a^{\dagger}$ ($a$) is the  quasi-particle creation (annihilation) operator which enter the RPA equation.  The occupation amplitudes ($u_{m\alpha}$ and $v_{m\alpha}$) satisfied the condition  $u^{2}_{m\alpha}$ + $v^{2}_{m\alpha}$ = 1 and were determined using the BCS equations.}

{The GT transitions were described with the help of the QRPA phonons}
\begin{equation}\label{co}
	A^{\dagger}_{\omega}(\mu)=\sum_{pn}[-Y^{pn}_{\omega}(\mu)a_{n}a_{\overline{p}} + X^{pn}_{\omega}(\mu)a^{\dagger}_{p}a^{\dagger}_{\overline{n}}].
\end{equation}
The indices $p$ and $n$ represent $m_{p}\alpha_{p}$ and $m_{n}\alpha_{n}$, respectively. {All neutron-proton pairs meeting the requirement $\pi_{n}.\pi_{p}$=1 (where $\pi$ denotes parity) and $\mu=m_{p}-m_{n}$= -1, 0, 1  were included in the summation. $X$ ($Y$) is the forward (backward) going amplitude and eigenfunction of the RPA equation.}  The energy eigenvalues was expressed by $\omega$. The pn-QRPA technique accounts for residual interaction between protons and neutrons via $p$-$p$ and $p$-$h$ channels, served by the interaction constants $\kappa$ and $\chi$, respectively. The $p$-$h$ GT force was calculated using
\begin{equation}\label{ph}
	V^{ph}= +2\chi\sum^{1}_{\mu= -1}(-1)^{\mu}Y_{\mu}Y^{\dagger}_{-\mu},
\end{equation}
with
\begin{equation}\label{y}
	Y_{\mu}= \sum_{j_{p}m_{p}j_{n}m_{n}}<j_{p}m_{p}\mid
	t_- ~\sigma_{\mu}\mid
	j_{n}m_{n}>c^{\dagger}_{j_{p}m_{p}}c_{j_{n}m_{n}}.
\end{equation}
In contrast, determination of $p$-$p$ GT interaction was conducted using
\begin{equation}\label{pp}
	V^{pp}= -2\kappa\sum^{1}_{\mu=
		-1}(-1)^{\mu}P^{\dagger}_{\mu}P_{-\mu},
\end{equation}
with
\begin{eqnarray}\label{p}
\begin{aligned}
		P^{\dagger}_{\mu}= \sum_{j_{p}m_{p}j_{n}m_{n}}<j_{n}m_{n}\mid
		(t_- \sigma_{\mu})^{\dagger}\mid
		j_{p}m_{p}>\\
		\times (-1)^{l_{n}+j_{n}-m_{n}}c^{\dagger}_{j_{p}m_{p}}c^{\dagger}_{j_{n}-m_{n}},
\end{aligned}
\end{eqnarray}
{where the operators would be defined below and remaining symbols have their usual meanings. The $\chi$ and $\kappa$ interaction constants were selected from Ref.~\cite{Homma}.   Reduced GT transition probabilities were obtained by transforming the QRPA ground state into one-phonon states in the daughter nucleus}
%\begin{equation}
%	B_{GT} (\omega) = |\langle \omega, \mu ||\tau_{\pm} \sigma_{\mu}||QRPA \rangle|^2.
%\end{equation}
\begin{equation}
	{	B_{GT} (\omega) = |\langle \omega, \mu ||\hat{O_\pm}||QRPA \rangle|^2,}
\end{equation}
{where  $ \hat{O}= \sum\limits_{i=1}^A\sum\limits_{\mu} \sigma_{\mu}(i)\tau_{\pm}(i) $,  $\sigma_{\mu}$ is the spin operator and $\tau_{\pm}$=$\tau_{x}\pm \iota \tau_{y}$  are the isospin raising and lowering operators, respectively. The model-independent Ikeda sum rule~\cite{Ikeda} can be evaluated for the operators using the relation}
\begin{equation}\label{ikeda}
	{	S_{-}+S_{+}= \sum\limits_{f}|\langle f |\hat{O_-}|i \rangle|^2+\sum\limits_{f}|\langle f |\hat{O_+}|i \rangle|^2=3(N-Z),}
\end{equation}
{where $|i \rangle $ and $|f \rangle $ are the initial and final states connected
by the GT operator, respectively. The Ikeda sum rule was found to be satisfied by our results. Details of our nuclear model can be found in ~\cite{Staudt,Hirsch,Muto}.}

\subsection{The RMF Model}
The RMF framework illustrates nuclei in a way that nucleons interact with one another by exchanging different mesons and photons~\cite{Walecka}. 
The introductory RMF model encountered  various difficulties in describing surface properties of nuclei and incompressibility of nuclear mater. For this reason, non linear version of the model was introduced (see Ref.~\cite{Ring} and references therein). Later counterpart version of the model such as density-dependent meson-exchange and point coupling were introduced and the model was referred to  as covariant density functional theory~\cite{Typel,Lalazissis,Meng,Nikic}. In the present investigation, we employed both the density-dependent meson-exchange framework and the point-coupling form of the model. In this subsection, we describe the density-dependent point-coupling versions of the model for an understanding.

In the density-dependent point-coupling version of the RMF model, isoscalar scalar $\sigma$ meson, the isoscalar vector $\omega$ meson, and the isovector vector $\rho$ meson fields are considered for the description of nuclear matter and single-particle nuclear properties. The starting point of density-dependent point-coupling version of the RMF model is a Lagrangian density given by

\begin{equation}
\begin{aligned}	
		\mathcal{L}  =\bar{\psi}(i\gamma.\partial -m)\psi 
		- \frac{1}{2}\alpha_S(\hat{\rho})(\bar{\psi}\psi)(\bar{\psi}\psi) - \frac{1}{2}\alpha_V(\hat{\rho})(\bar{\psi}\gamma^\mu\psi)\\(\bar{\psi}\gamma_\mu\psi) - \frac{1}{2}\alpha_{TV}(\hat{\rho})(\bar{\psi}\overrightarrow{\tau}\gamma^\mu\psi)(\bar{\psi}\overrightarrow{\tau}\gamma_\mu\psi)
		- \frac{1}{2}\delta_S(\partial_\nu\bar{\psi}\psi)\\ (\partial^\nu\bar{\psi}\psi) - e\bar{\psi}.A\frac{(1-\tau_3)}{2}\psi.	
		\label{lagden}
\end{aligned}
\end{equation}
Eq.~(\ref{lagden}) covers terms such as the free-nucleon Lagrangian, the point-coupling interaction terms and the coupling of the protons to the electromagnetic field. The single-nucleon Dirac equation with nucleon self-energies may be obtained from the variation of the Lagrangian with respect to $\bar{\psi}$. Details and discussion about functional form of the couplings can be found in Ref.~\cite{Nikic}.

In the present study, we used the DIRHB code~\cite{NikicT} to obtain PECs of ($^{24}$Mg, $^{28}$Si, and $^{32}$S) nuclei as a function of $\beta_2$ using both density-dependent meson-exchange DD-ME2~\cite{Lalazissis} and point-coupling DD-PC1~\cite{Nikic} functionals. For this purpose quadrupole moment constrained RMF calculations were performed. Pairing correlations play important role for open-shell nuclei and the BCS approximation was used to tackle these correlations. Furthermore, constant G approximation~\cite{Karatzikos} was used for the PEC calculations.
%%%%%%%%%%%%%%%%%%%%%%%%%%%%%%%%%%%%%%%%%%%%%%%%%%%%%%%%
%%%%%%%%%% RESULTS AND DISCUSSIONS %%%%%%%%%%%%%%%
\section{Results and Discussion}
In this particular analysis, we employed the $\beta_2$ values, acquired by the FRDM and RMF approaches, as a free {parameter in the pn-QRPA model to investigate the GT strength of the selected nuclei. We first determine the ground-state deformation variables of these nuclei. After reviewing the characteristics of nuclear ground state properties, we continue to discuss how GT transitions were computed employing the pn-QRPA model.}
\begin{table}[h!]
	\centering
	\caption{ $\beta_2$ values adopted from the FRDM \cite{Moller} and those computed via RMF framework with DD-ME2 and DD-PC1 interactions. }
	\label{tab:1}
	\addtolength{\tabcolsep}{1pt}
	\scalebox{1.0}{
		\begin{tabular}{@{}cccc@{}}
			\hline\noalign{\smallskip}
			\multicolumn{1}{c}{Nuclei }&\multicolumn{3}{c}{$\beta_2$ }\\
			\hline
			& FRDM& DD-ME2 & DD-PC1  \\\hline
			$^{24}$Mg& 0.393 &   $0.508$  &  0.525 \\	
			$^{28}$Si & -0.363&   -0.377  &  -0.376 \\	
			%$^{28}$Si	& &   $0.032$  &  0.096 \\\hline
			$^{32}$S& 0.221 &   $0.263$  &  0.243 \\\hline		
	\end{tabular}}
\end{table}
\begin{figure*}[h!]
	\centering
	{\includegraphics[width=.95\textwidth]{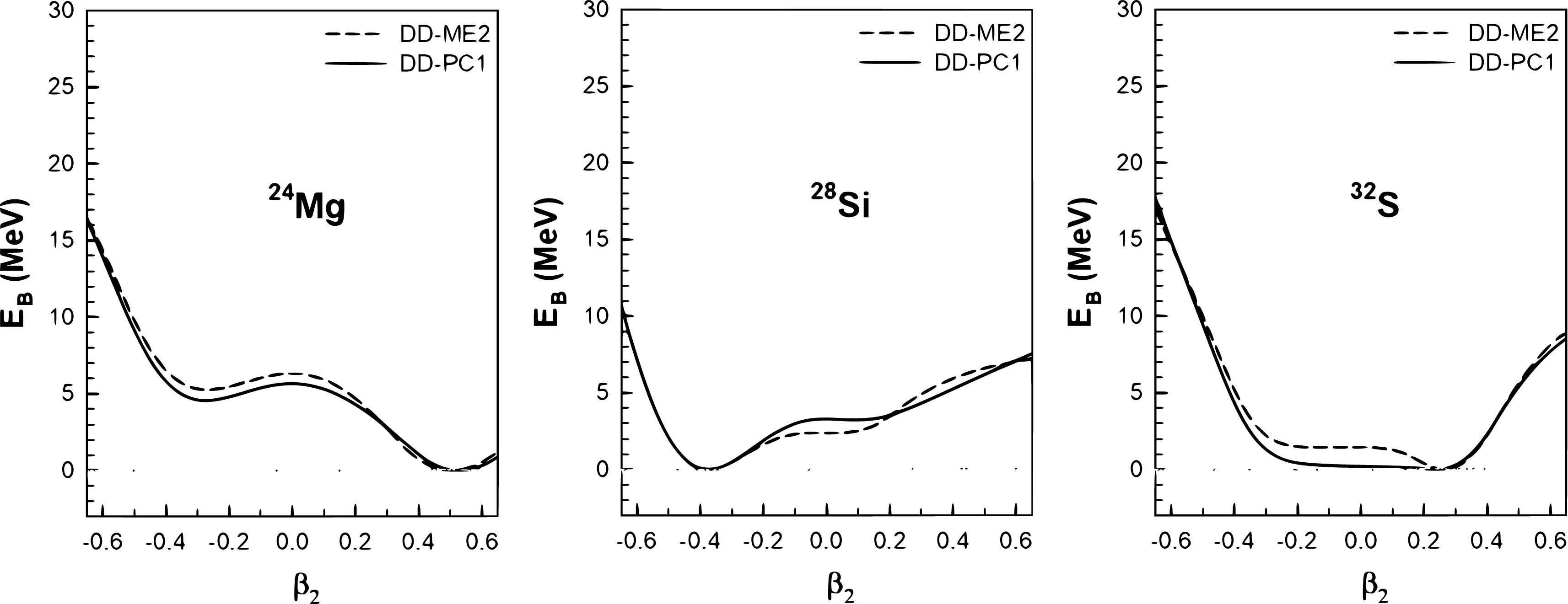}}
	%	\vspace*{-60mm}
	\caption{Calculated PECs of  $^{24}$Mg, $^{28}$Si and $^{32}$S  using the RMF model with DD-ME2 and DD-PC1 functionals.}
	\label{fig:0}
\end{figure*}

One of the most successful nuclear models to study deformation of nuclei is the RMF model~\cite{Ring}. In this study, we have employed axially symmetric RMF model calculation , using two density-dependent versions,  for obtaining the ground-state quadrupole moment deformation values of the $N$=$Z$ ($^{24}$Mg, $^{28}$Si, and $^{32}$S) nuclei. In our calculation, the density-dependent meson-exchange DD-ME2~\cite{Lalazissis} and point-coupling DD-PC1~\cite{Nikic} functionals were used. For prediction of the  ground-state deformation values of nuclei, we made use of the potential energy curves (PECs). For this purpose, we have employed quadrupole moment constrained axially symmetric RMF model calculation as prescribed in Ref.~\cite{Nikic}. We obtained binding energies of the selected nuclei as a function of deformation parameter $\beta_2$. In Fig.~\ref{fig:0} the calculated PECs of $^{24}$Mg, $^{28}$Si, and $^{32}$S are shown. It should be noted that energy for the ground-state was taken as a reference in each curve. As seen from Fig.~\ref{fig:0}, both the DD-ME2 and DD-PC1 functionals produce similar PECs {for the selected $sd$-shell nuclei}. The RMF model predicts high deformation values for $^{24}$Mg and $^{28}$Si. The ground-state shape of $^{24}$Mg and $^{28}$Si is expected to be prolate and oblate, respectively. For the case of $^{32}$S, DD-ME2 indicates prolate shape while DD-PC1 shows almost flat PEC. The most probable shape seems to be prolate for $^{32}$S. The computed ground-state deformation values of the $sd$-shell nuclei are listed in Table~\ref{tab:1}. 

The $\beta_2$ values were used  as a free parameter in our investigation. Our preliminary calculation for the GT strength were based on the $\beta_2$ values adopted from the FRDM~\cite{Moller} and represented as {pn-QRPA-0}. The $\beta_2$ values computed by the RMF model using DD-PC1 and DD-ME2 interactions were used as the foundation for the second and third sets of calculation. These calculations are referred to as {pn-QRPA-1} and {pn-QRPA-2}, respectively. 
\begin{figure}[h!]
	\centering
	{\includegraphics[width=1.400\textwidth]{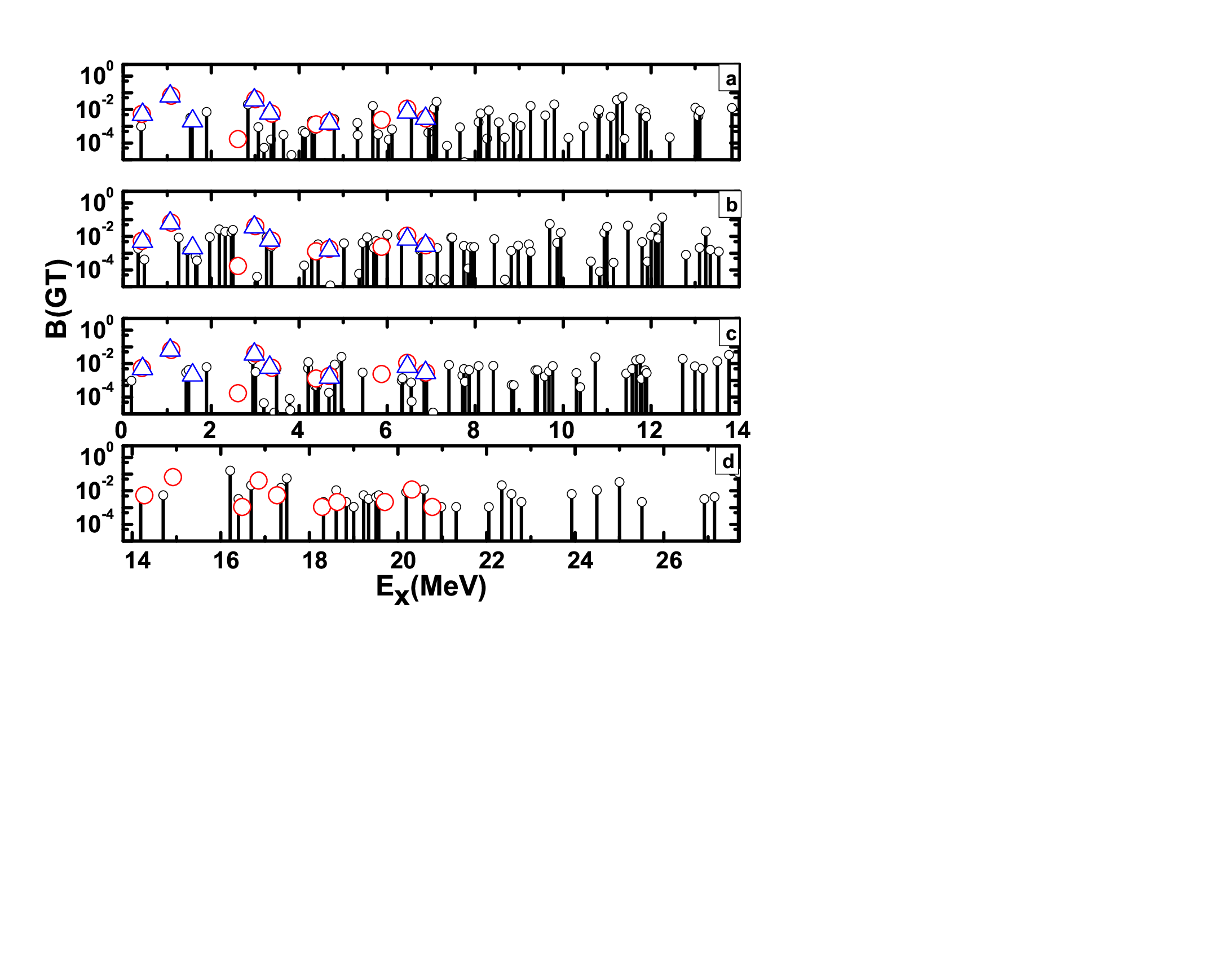}}
	\vspace*{-60mm}
	\caption{Computed GT strength distributions of $^{24}$Mg. (a) {pn-QRPA-0}. (b)  {pn-QRPA-1}. (c)  {pn-QRPA-2}.  (d) The calculated GT strength distributions of Ref.~\cite{Eunja}. The measured data (triangles) and (big circles) were taken from~\cite{Anderson} and \cite{Zegers}, respectively.}
	\label{fig:1}
\end{figure}
{ The calculated GT strengths exhaust almost 100\% of the sum rule (\ref{ikeda}) up to $E_x$=35~MeV.  In the present cases of $N$=$Z$ nuclei, equal GT strength was calculated in the $\tau_{+}$ and $\tau_{-}$ channels.} The present calculations are shown in Fig.~\ref{fig:1} together with the experimental results from Refs.~\cite{Anderson,Zegers} up to $E_\mathrm{x}$=12 MeV.   We aim to explore the effect  of $\beta_2$, computed by various interactions, on the GT strength distributions. The values of $\beta_2$ used in our analysis are shown in Table~\ref{tab:1}. In the present investigation, modest differences in $\beta_2$ values {affected the calculated GT distributions, albeit marginally}.  In general, a satisfactory agreement was found between the GT strength of our model based results and measurements \cite{Anderson,Zegers}. But there are a few discrepancies at low excitation energies.  {The pn-QRPA calculation based on the RMF interactions  predicted the strength distribution better than the {pn-QRPA-0} interaction.}  The splitting of the GT strength into a strong and weak state close to 2.8 MeV fits well with the experimental results, as illustrated in Figs.~(\ref{fig:1})b-(\ref{fig:1})c.  The {pn-QRPA-0 interaction performs marginally better} in predicting the location of individual states at high excitation energies, as depicted in Fig.~(\ref{fig:1})a. Furthermore, we have displayed the computed results of Ref. \cite{Eunja} and comparison with measured data~\cite{Zegers} in Fig.~(\ref{fig:1})d. A point to be emphasized is that in the investigation of Ref. \cite{Eunja}, the authors presented their results with respect to the excitation energy from the parent nucleus. 
\begin{figure}[h!]
	\centering
	{\includegraphics[width=1.40\textwidth]{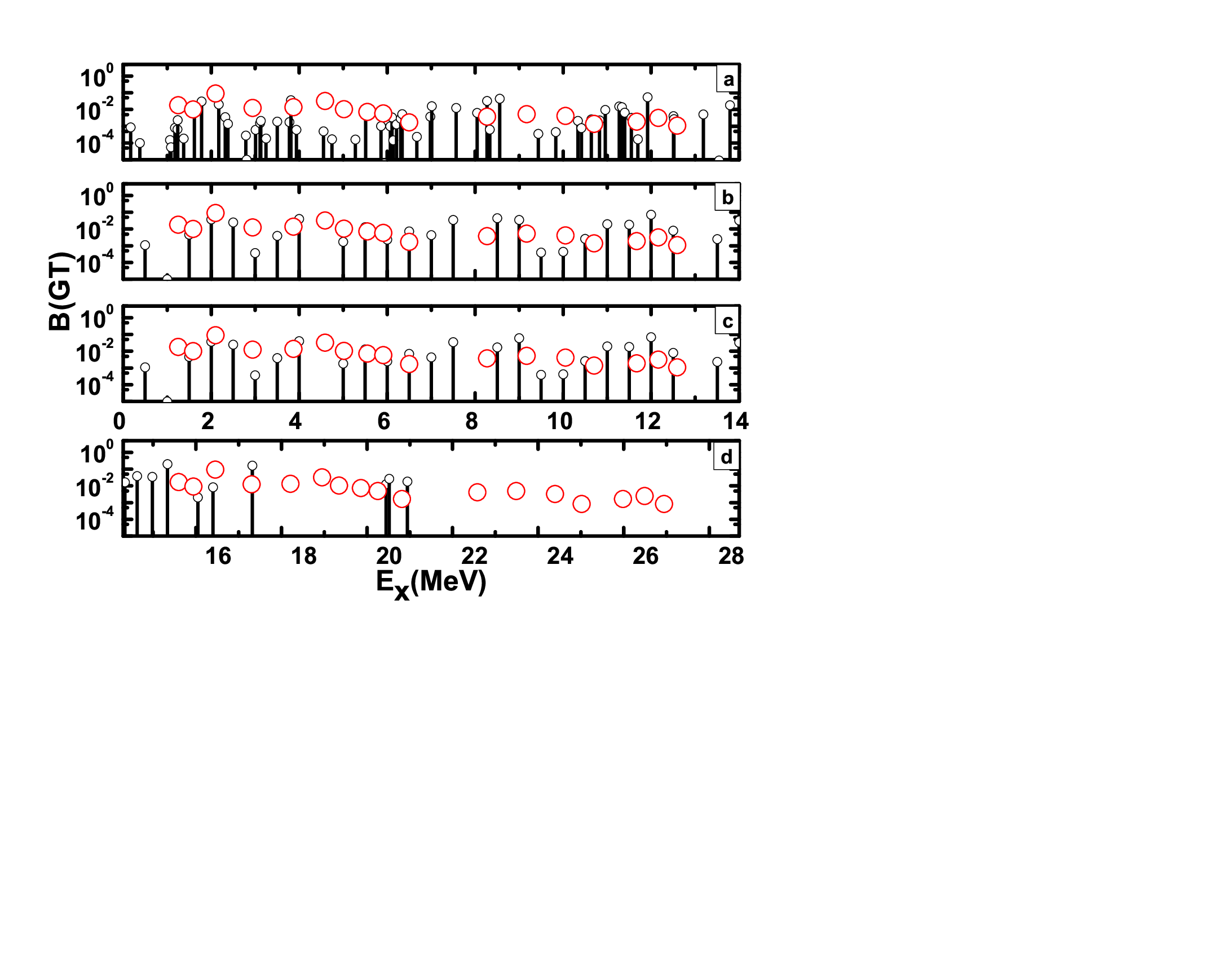}}
	\vspace*{-60mm}
	\caption{Computed GT strength distributions of $^{28}$Si. (a) {pn-QRPA-0}. (b)  {pn-QRPA-1}. (c)  {pn-QRPA-2}. (d) The calculated GT strength distributions of Ref.~\cite{Eunja}.  Experimental data (big circles) was taken from Ref.~\cite{Anderson}.}
	\label{fig:2}
\end{figure}

\begin{figure}[h!]
	\centering
	{\includegraphics[width=1.40\textwidth]{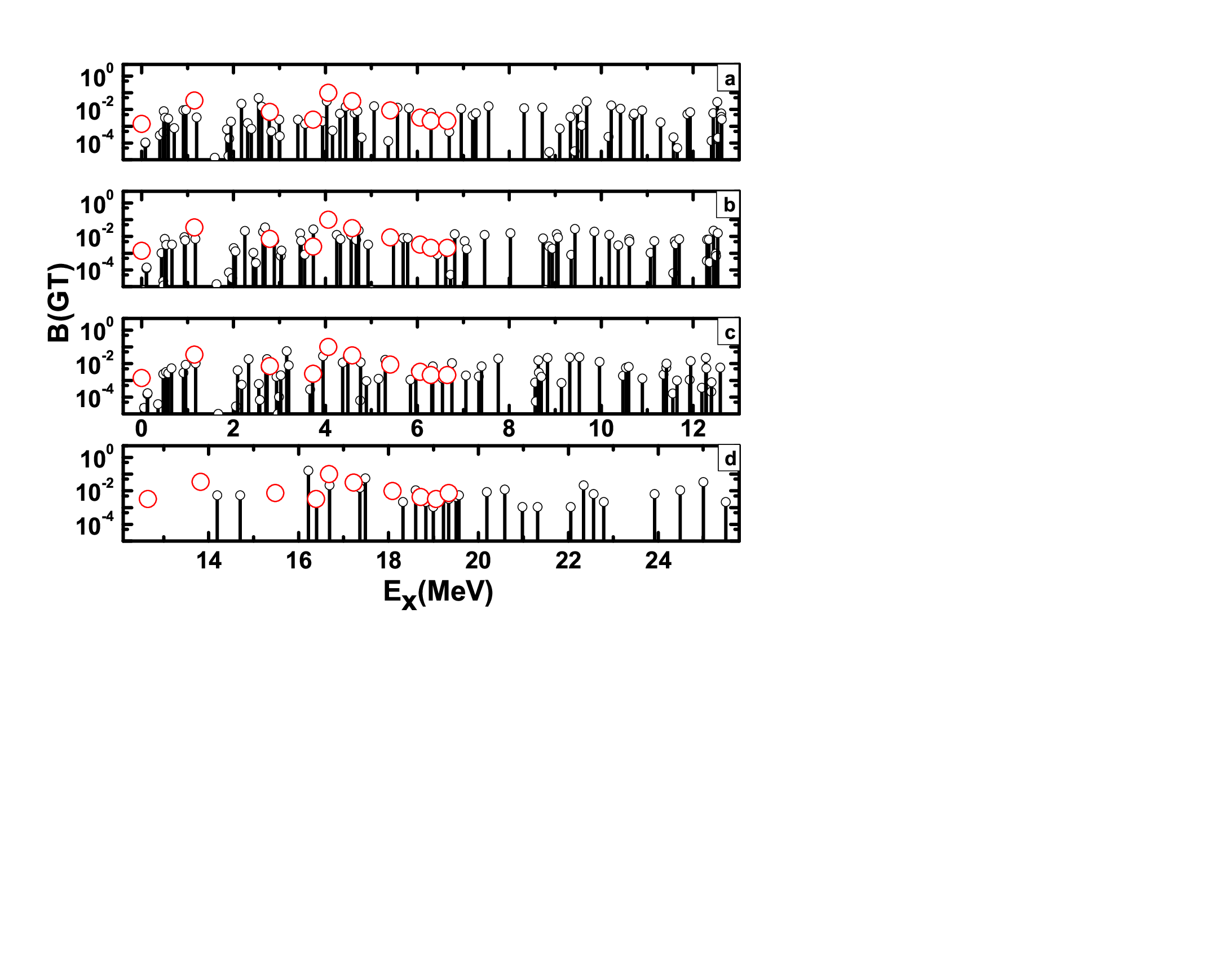}}
	\vspace*{-60mm}
	\caption{Computed GT strength distributions of $^{32}$S. (a)  {pn-QRPA-0}. (b)  {pn-QRPA-1}. (c)  {pn-QRPA-2}. (d) The computed GT strength distributions of Ref.~\cite{Eunja}. The measured data (big circles) was taken from~\cite{Grewe}.}
	\label{fig:3}
\end{figure}

Fig.~\ref{fig:2} compares the calculated and experimental B(GT) around $E_\mathrm{x}$=14~MeV for the oblate shape of $^{28}$Si. We performed the pn-QRPA calculation using three different interactions, as discussed before.  It can be seen that the pn-QRPA model predicted well-fragmented GT strength distributions for $^{28}$Si. The {pn-QRPA-0} interaction produced more fragmentation of the GT strength as shown in Fig.~(\ref{fig:2})a. The {pn-QRPA-1} and {pn-QRPA-2} predict relatively stronger peaks as depicted in Figs.~(\ref{fig:2})b-(\ref{fig:2})c.    For the $A$=28 nucleus, the comparison between measured and pn-QRPA predicted GT distributions is better both qualitatively and quantitatively. The states splitting were identified within $E_\mathrm{x}$=(0--14)~MeV by employing both the {pn-QRPA-0} and pn-QRPA-RMF interactions. Fig.~(\ref{fig:2})d represents the comparison between the model based results of Ref.~\cite{Anderson} and Ref. \cite{Eunja}. One can see that states do not exist at higher energies except near 20~MeV in Ref.~\cite{Eunja} (it is reminded that authors presented their results with respect to the excitation energy from the parent nucleus). However, our investigations cover the whole range of the experimental study. It is obvious that our present model based results are in good comparison with the measured data~\cite{Anderson}. The state splitting appeared above $E_\mathrm{x}$=8~MeV using the {pn-QRPA-1} interaction and below  $E_\mathrm{x}$=7~MeV using the {pn-QRPA-2} interaction, although splitting are comparatively weaker in the {pn-QRPA-2} interaction.

%In Fig.~(\ref{fig:20})c the states do not exist at higher energies except near 20~MeV. However, our investigation covers the whole range of the experimental study, both qualitatively and quantitatively.

Fig.~\ref{fig:3} shows the comparison between our model-based results and the measured GT strength of \cite{Grewe}. The GT strength in Fig.~\ref{fig:3} is appropriately distributed between low and high excitations energies displaying a slight variation with $\beta_2$ values.  Fig.~(\ref{fig:3})a shows the comparison between the {pn-QRPA-0} and measured data \cite{Grewe}. Figs.~(\ref{fig:3})b-(\ref{fig:3})c depict a similar comparison of {pn-QRPA-1} and {pn-QRPA-2} interaction with experimental data,  respectively. Fig.~(\ref{fig:3})d displays the model based results of Ref.~\cite{Eunja} and comparison with measured data~\cite{Grewe} from the parent state excitation energies.  Among the {pn-QRPA-0}, {pn-QRPA-1}, and {pn-QRPA-2} interactions, the {pn-QRPA-2}   is best-fitted with measured data for $^{32}$S. 

\section{Conclusion}
The present investigations cover the nuclear structure properties of selected $sd$-shell nuclei,  $^{24}$Mg, $^{28}$Si, and $^{32}$S. We calculated the nuclear ground state deformation parameter for $N$=$Z$ nuclei employing the axially deformed RMF framework, utilizing two distinct types of density dependent interactions.
The pn-QRPA framework with a deformed basis was later used to investigate the effect of $\beta_2$ values on predicted GT distributions. It became apparent that the predicted strength distributions were altered with the $\beta_2$ values, albeit marginally.  Within the range of measured energies, the predicted GT strength distributions agreed well with the experimental data. The {pn-QRPA-2} interaction resulted in the best fit for the measured distribution.

\section*{Acknowledgements}
The authors would like to acknowledge the useful discussion with Mr. Muhammad Tahir. J.-U. Nabi would like to acknowledge the support of the Higher Education Commission Pakistan through project
20-15394/NRPU/R\&D/HEC/2021.

\end{document}